\documentclass[twocolumn]{autart}

\usepackage{graphicx}
\usepackage{amsmath,amssymb}

\newtheorem{theorem}[thm]{Theorem}

\newtheorem{proposition}[thm]{Proposition}

\newtheorem{definition}[thm]{Definition}
\newtheorem{remark}[thm]{Remark}
\newtheorem{example}[thm]{Example}
\newenvironment{proof}{\pf}{\endpf}

\newcommand{\del}{\partial}
\newcommand{\pdiff}[2]{\frac{\del #1}{\del #2}}

\newcommand{\bbR}{\mathbb{R}}
\newcommand{\bbZ}{\mathbb{Z}}

\newcommand{\bbT}{\mathbb{T}}

\newcommand{\st}{\mathrm{subject\ to\ }}

\newcommand{\rmstate}{\mathrm{state}}
\newcommand{\rminput}{\mathrm{input}}
\newcommand{\rmzero}{\mathrm{zero}}
\newcommand{\rmc}{\mathrm{c}}
\newcommand{\rmd}{\mathrm{d}}
\newcommand{\rme}{\mathrm{e}}
\newcommand{\soft}{\mathrm{soft}}
\newcommand{\req}{\mathrm{req}}

\newcommand{\rmp}{\mathrm{p}}

\newcommand{\eng}{\mathrm{eng}}
\newcommand{\mot}{\mathrm{mot}}
\newcommand{\brk}{\mathrm{brk}}
\newcommand{\cmd}{\mathrm{cmd}}
\newcommand{\rmds}{\mathrm{ds}}
\newcommand{\soc}{\mathrm{SoC}}
\newcommand{\target}{\mathrm{target}}
\newcommand{\rmref}{\mathrm{ref}}
\newcommand{\reg}{\mathrm{reg}}

\begin{document}

\begin{frontmatter}

\title{Modeling and Control of Deep Sign-Definite Dynamics\\ with Application to Hybrid Powertrain Control}

\thanks[footnoteinfo]{This paper was not presented at any IFAC meeting. Corresponding author T.~Kato.}

\author[TCRDL]{Teruki Kato}\ead{teruki.kato.tg@mosk.tytlabs.co.jp},
\author[TCRDL]{Ryotaro Shima}\ead{ryotaro.shima@mosk.tytlabs.co.jp},
\author[KU]{Kenji Kashima}\ead{kk@i.kyoto-u.ac.jp}

\address[TCRDL]{Toyota Central R\&D Labs., Inc., 41-1 Yokomichi, Nagakute, Aichi, Japan.}
\address[KU]{Graduate School of Informatics, Kyoto University, Yoshida-honmachi, Sakyo-ku, Kyoto, Japan}


\begin{abstract}
Data-driven control increasingly relies on deep models for complex systems whose first-principles models are difficult to obtain.
For reliable deployment, however, learned dynamics should respect physical structure and lead to tractable optimal control.
We introduce sign constraints, namely sign restrictions on Jacobian entries, as a unified description of monotonicity, positivity, and sign-definiteness.
For exactly linearizable deep dynamics, we provide structural conditions and neural-network parameterizations that enforce these constraints by construction.
The same structure also allows model predictive control to be formulated as a convex quadratic program or as a convex relaxation, yielding a unique optimizer and a Lipschitz continuous control law.
Applications to a three-tank system and a hybrid powertrain demonstrate that the proposed approach offers improved extrapolation performance and smoother control inputs compared with competing nonconvex formulations.
\end{abstract}

\end{frontmatter}

\section{Introduction}
Data-driven control is becoming essential for complex engineering systems whose dynamics are costly to derive from first principles \cite{brunton_kutz}.
Deep neural networks (NNs) are attractive because they can approximate nonlinear dynamics from data, but control-oriented learning also requires physical consistency and reliable optimization.
Without structural restrictions, a learned model may contradict known cause-effect directions: for example, in a tank system, increasing pump voltage should not decrease the downstream water level.
Such knowledge can be expressed as sign restrictions on Jacobian entries.
Monotonicity \cite{monotone_control_systems}, positivity \cite{farina_positive}, and sign-definiteness \cite{sign_def_fb} are important special cases.

Physics-informed machine learning (PIML) has been studied extensively \cite{brunton_kutz}.
A common PIML strategy is to add physical penalties to the training loss, but this does not necessarily guarantee the property after training.
Input convex neural networks (ICNNs) \cite{icnn} can structurally impose nonnegativity of weights and hence monotonicity, yet a standard quadratic tracking objective with constraints generally remains nonconvex.
Exactly linearizable (EL) deep models \cite{exlin} learn nonlinear state and input transformations together with linear dynamics in transformed coordinates, which enables convex unconstrained regulation.
However, a general framework for sign constraints and constrained tracking has not been established.
As with EL models, Koopman-based deep models with state-dependent input transformations \cite{brunton_kutz,fan_stable_koopman} lack a general framework for sign constraints, and constrained tracking can remain nonconvex.

This paper targets systems for which at least part of the Jacobian sign information is known from physics; unknown signs can be left unconstrained.
The contributions are as follows.
\begin{itemize}
    \item We formulate sign constraints that unify monotonicity, positivity, and sign-definiteness.
    \item We incorporate sign constraints into EL models, improving physical plausibility and extrapolation.
    \item We formulate constrained tracking model predictive control (MPC) \cite{mpc_handbook} as a convex quadratic program (QP) or as a convex relaxation, improving continuity and computational reliability.
    \item We demonstrate the method on a three-tank benchmark \cite{sysid_benchmark} and a hybrid powertrain \cite{vehicle_propulsion_systems}.
\end{itemize}
Table~\ref{tab:method_comparison} summarizes the relationship with representative model classes.
\begin{table}[t]
    \centering
    \caption{Comparison of learned dynamics models for control.}
    \renewcommand{\arraystretch}{1.18}
    \setlength{\tabcolsep}{1.5pt}
    \scriptsize
    \begin{tabular}{|c|c|c|c|}
        \hline
        Model & Sign & Regulation & Constrained Tracking \\
        \hline
        NN & None & Nonconvex & Nonconvex \\
        \hline
        ICNN & Monotonicity & Nonconvex & Nonconvex \\
        \hline
        EL & None & Convex & Nonconvex \\
        \hline
        Proposed EL & General & Convex & Convex/relaxed \\
        \hline
    \end{tabular}
    \label{tab:method_comparison}
\end{table}

The notation is as follows.
We denote by $\bbR^n_\geq$ the set of all $n$-dimensional nonnegative real vectors and by $\bbZ_\geq$ the set of all nonnegative integers.
For a positive integer $N$, $\bbZ_N:=\{1,\ldots,N\}$.
The order $\leq$ is componentwise.
For $r\in\bbR$, $\sigma(r)=1$ if $r>0$, $\sigma(r)=-1$ if $r<0$, and $\sigma(r)=0$ if $r=0$.
We write $\operatorname{ReLU}(r):=\max\{r,0\}$.
Operator $\odot$ denotes the componentwise product, and $\delta_{ij}$ is the Kronecker delta.
For a continuous-time system, $x^+(t):=\dot{x}(t)$ and $\bbT_\geq:=\bbR_\geq$; for a discrete-time system, $x^+(t):=x(t+1)$ and $\bbT_\geq:=\bbZ_\geq$.
A map $T:\bbR^n\to\bbR^n$ is componentwise strictly increasing if $T_i(z)=T_i(z_i)$ and every scalar function $T_i$ is strictly increasing.

\section{Standard Exactly Linearizable Models \cite{exlin}}
This section reviews standard EL models.
They are not local Jacobian linearizations; rather, they restrict the learned model class so that the nonlinear dynamics become exactly linear after a nonlinear change of coordinates.
Throughout this paper, the state is assumed to be directly observable.

\begin{figure}[t]
    \centering
    \includegraphics[width=0.99\linewidth]{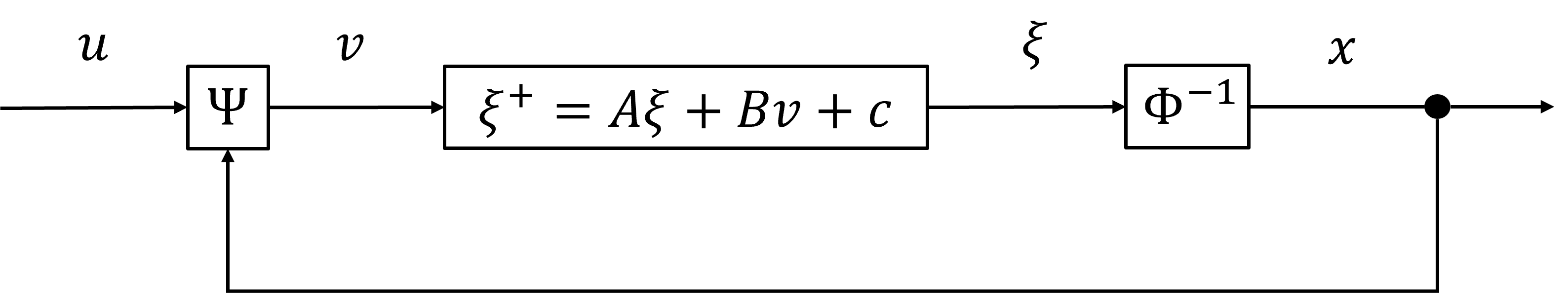}
    \caption{Exactly linearizable model.}
    \label{fig:exlin}
\end{figure}

Consider the continuous- and discrete-time EL models
\begin{align}
    &\dot{x}(t)=\left.\pdiff{\Phi}{x}\right|_{x(t)}^{-1}
    \big[A\Phi(x(t))+B\Psi(u(t);x(t))+c\big], \label{eq:standard_el_ct}\\
    &x(t+1)=\Phi^{-1}\big(A\Phi(x(t))+B\Psi(u(t);x(t))+c\big). \label{eq:standard_el_dt}
\end{align}
Fig.~\ref{fig:exlin} shows the EL model structure.
We define the transformed state and input coordinates by $\xi:=\Phi(x)$ and $v:=\Psi(u;x)$.
The state transformation $\Phi:\bbR^{n_x}\to\bbR^{n_x}$ is a diffeomorphism in $x$, and $\Psi:\bbR^{n_u}\times\bbR^{n_x}\to\bbR^{n_u}$ is, for each fixed $x$, a diffeomorphism in $u$.
The transformed dynamics satisfy
\begin{align}
    \xi^+(t)=A\xi(t)+Bv(t)+c . \label{eq:transformed_linear}
\end{align}
The matrices $A,B$, vector $c$, and transformations $\Phi,\Psi$ are learned jointly from data $\{x(t),u(t)\}_{t=0}^T$ by minimizing, for the discrete-time case,
\begin{align}
    L(A,B,c,\Phi,\Psi)&=\frac{1}{T}\sum_{t=0}^{T-1}\Big\|
    x(t+1)-\Phi^{-1}\big(\eta(t)\big)\Big\|^2, \label{eq:loss_exlin}\\
    \eta(t)&:=A\Phi(x(t))+B\Psi(u(t);x(t))+c. \nonumber
\end{align}
\begin{remark}[Equilibrium preservation]\label{rem:equilibrium_preservation}
If a known equilibrium $(x_\rme,u_\rme)$ should be preserved, define $\xi_\rme:=\Phi(x_\rme)$ and $v_\rme:=\Psi(u_\rme;x_\rme)$.
Then one can set $c=-A\xi_\rme-Bv_\rme$ in continuous time and $c=\xi_\rme-A\xi_\rme-Bv_\rme$ in discrete time instead of learning $c$ freely.
\end{remark}

For unconstrained regulation to $(x_\rme,u_\rme)$, the formulation in transformed coordinates
\begin{align}
    &\min_{v}\sum_{t=0}^{N-1} w^\xi\|\xi(t)-\xi_\rme\|^2
    + w^v\|v(t)-v_\rme\|^2 \nonumber\\
    &\st\quad \xi(t+1)=A\xi(t)+Bv(t)+c \label{eq:standard_el_regulation} \\
    &\hspace{30mm} t=0,\ldots,N-1 \nonumber
\end{align}
with $w^\xi, w^v>0$ is a convex QP. 
This is a principal advantage of EL models.
However, without additional structure, the signs of the Jacobian of the original dynamics cannot be imposed simply from the signs of $A,B,c$.
Moreover, original-coordinate state and input bounds in constrained tracking generally become nonconvex after transformation.
The remaining sections introduce a structure that addresses these two limitations.

\section{Existing Sign-Related Properties \cite{monotone_control_systems,farina_positive,sign_def_fb}}
Consider nonlinear system
\begin{align}
    x^+(t)=f(x(t),u(t)) \quad (t\in\bbT_\geq) \label{eq:state_eq_nonlin}
\end{align}
with $x(t)\in\bbR^{n_x}$, $u(t)\in\bbR^{n_u}$, and $f:\bbR^{n_x}\times\bbR^{n_u}\to\bbR^{n_x}$ of class $C^1$.
In continuous time, we denote the vector field by $f^\rmc$, and in discrete time, we denote the update map by $f^\rmd$.

\begin{definition}[Monotonicity]\label{def:monotone}
Let $x(t)$ be the trajectory of \eqref{eq:state_eq_nonlin} corresponding to initial state $x(0)$ and input $u(t)$, and let $x'(t)$ correspond to $x'(0)$ and $u'(t)$.
System \eqref{eq:state_eq_nonlin} is monotone if $x(0)\leq x'(0)$ and $u(t)\leq u'(t)$ for all $t\in\bbT_\geq$ imply $x(t)\leq x'(t)$ for all $t\in\bbT_\geq$.
\end{definition}
\begin{proposition}[Kamke conditions]\label{prop:nonlin_monotone_jacobi}
System \eqref{eq:state_eq_nonlin} is monotone if and only if, for all $x\in\bbR^{n_x}$ and $u\in\bbR^{n_u}$,
$\pdiff{f_i}{x_j}(x,u)\geq0$ for all $i,j\in\bbZ_{n_x}$ with $i\neq j$ when $f=f^\rmc$, and $\pdiff{f_i}{u_j}(x,u)\geq0$ for all $i\in\bbZ_{n_x}$ and $j\in\bbZ_{n_u}$.
\end{proposition}
\begin{definition}[Positivity]
System \eqref{eq:state_eq_nonlin} is positive if $x(0)\geq0$ and $u(t)\geq0$ for all $t\in\bbT_\geq$ imply $x(t)\geq0$ for all $t\in\bbT_\geq$.
\end{definition}
\begin{proposition}\label{prop:nonlin_pos_f_cont}
For continuous-time system \eqref{eq:state_eq_nonlin}, positivity holds if and only if, for any $x\in\bbR^{n_x}_\geq$, $u\in\bbR^{n_u}_\geq$, and $i\in\bbZ_{n_x}$,
\begin{align}
    f_i^\rmc(x_1,\ldots,x_{i-1},0,x_{i+1},\ldots,x_{n_x},u)\geq0 .
\end{align}
\end{proposition}
\begin{proposition}\label{prop:nonlin_pos_f_disc}
For discrete-time system \eqref{eq:state_eq_nonlin}, positivity holds if and only if $f^\rmd(x,u)\geq0$ for all $x\in\bbR^{n_x}_\geq$ and $u\in\bbR^{n_u}_\geq$.
\end{proposition}
\begin{proposition}\label{prop:nonlin_monotone_pos}
If \eqref{eq:state_eq_nonlin} is monotone and $f(0,0)\geq0$, then \eqref{eq:state_eq_nonlin} is positive.
\end{proposition}
\begin{definition}[Sign-definiteness]\label{def:sign_def}
System \eqref{eq:state_eq_nonlin} has sign-definiteness determined by signs $s_{ij}^\rmstate\in\{1,-1,0\}$ for $i,j\in\bbZ_{n_x}$ with $i\neq j$ when $f=f^\rmc$, and $s_{ij}^\rminput\in\{1,-1,0\}$ for $i\in\bbZ_{n_x}$, $j\in\bbZ_{n_u}$ if, for all $x\in\bbR^{n_x}$ and $u\in\bbR^{n_u}$,
\begin{align}
    \sigma\!\left(\pdiff{f_i}{x_j}(x,u)\right)&=s_{ij}^\rmstate, \\
    \sigma\!\left(\pdiff{f_i}{u_j}(x,u)\right)&=s_{ij}^\rminput .
\end{align}
\end{definition}

\section{Sign Constraints}
We now introduce sign constraints that generalize existing sign-related properties.

\subsection{Sign constraints for general nonlinear systems}
\begin{definition}[Sign constraints]\label{def:sign_constraint}
System \eqref{eq:state_eq_nonlin} satisfies sign constraints determined by sign sets $S_{ij}^\rmstate\subset\{1,-1,0\}$ for $i,j\in\bbZ_{n_x}$ with $i\neq j$ when $f=f^\rmc$, and $S_{ij}^\rminput\subset\{1,-1,0\}$ for $i\in\bbZ_{n_x}$, $j\in\bbZ_{n_u}$ if, for all $x\in\bbR^{n_x}$ and $u\in\bbR^{n_u}$,
\begin{align}
    \sigma\!\left(\pdiff{f_i}{x_j}(x,u)\right)&\in S_{ij}^\rmstate, \label{eq:sign_state_general}\\
    \sigma\!\left(\pdiff{f_i}{u_j}(x,u)\right)&\in S_{ij}^\rminput. \label{eq:sign_input_general}
\end{align}
Moreover, \eqref{eq:state_eq_nonlin} satisfies sign constraints at the origin determined by $S_i^\rmzero\subset\{1,-1,0\}$ for $i\in\bbZ_{n_x}$ if
\begin{align}
    \sigma(f_i(0,0))\in S_i^\rmzero . \label{eq:sign_origin_general}
\end{align}
\end{definition}
\begin{example}
If $S_{ij}^\rmstate=\{s_{ij}^\rmstate\}$ and $S_{ij}^\rminput=\{s_{ij}^\rminput\}$, Definition~\ref{def:sign_constraint} reduces to sign-definiteness.
If $S_{ij}^\rmstate=S_{ij}^\rminput=\{1,0\}$, it enforces monotonicity.
Adding $S_i^\rmzero=\{1,0\}$ yields a sufficient condition for positivity by Proposition~\ref{prop:nonlin_monotone_pos}; for the linear and EL classes below, this condition is also necessary.
The choice $\{1,-1,0\}$ represents no constraint.
\end{example}
\begin{remark}
Definition~\ref{def:sign_constraint} imposes the same sign condition for every $x\in\bbR^{n_x},u\in\bbR^{n_u}$.
The definition can be restricted to a specified subset of the state-input space when only local sign knowledge is available.
\end{remark}
\subsection{Continuous- and discrete-time relations}
For the Euler discretization $f^\rmd(x,u):=x+\Delta t f^\rmc(x,u)$ with $\Delta t>0$, off-diagonal state signs, input signs, and origin signs are preserved from $f^\rmc$ to $f^\rmd$:
\begin{align}
    &\pdiff{f_i^\rmd}{x_j}=\delta_{ij}+\Delta t\pdiff{f_i^\rmc}{x_j},\quad
    \pdiff{f_i^\rmd}{u_j}=\Delta t\pdiff{f_i^\rmc}{u_j},\nonumber\\
    &f_i^\rmd(0,0)=\Delta t f_i^\rmc(0,0). \label{eq:ct_dt_sign_relation}
\end{align}
Diagonal state signs are not part of the continuous-time definition and hence are not included in this preservation statement.
Conversely, if $f^\rmd$ satisfies sign constraints, the continuous-time approximation $f^\rmc(x,u):=(f^\rmd(x,u)-x)/\Delta t$ preserves all signs because
\begin{align}
    \pdiff{f_i^\rmc}{x_j}&=\frac{1}{\Delta t}\left(\pdiff{f_i^\rmd}{x_j}-\delta_{ij}\right),\nonumber\\
    \pdiff{f_i^\rmc}{u_j}&=\frac{1}{\Delta t}\pdiff{f_i^\rmd}{u_j},\quad
    f_i^\rmc(0,0)=\frac{f_i^\rmd(0,0)}{\Delta t}. \label{eq:dt_ct_sign_relation}
\end{align}
\subsection{Sign constraints for linear systems}
For the linear system \eqref{eq:transformed_linear}, sign constraints are characterized entrywise.
\begin{proposition}\label{prop:linear_sign}
System \eqref{eq:transformed_linear} satisfies \eqref{eq:sign_state_general}--\eqref{eq:sign_origin_general} if and only if
\begin{align}
    \sigma(A_{ij})&\in S_{ij}^\rmstate &&(i,j\in\bbZ_{n_x},\ i\neq j\ \text{if } f=f^\rmc), \label{eq:sign_cond_A}\\
    \sigma(B_{ij})&\in S_{ij}^\rminput &&(i\in\bbZ_{n_x},\ j\in\bbZ_{n_u}), \label{eq:sign_cond_B}\\
    \sigma(c_i)&\in S_i^\rmzero &&(i\in\bbZ_{n_x}). \label{eq:sign_cond_c}
\end{align}
\end{proposition}
\begin{proof}
Let $f(x,u):=Ax+Bu+c$. Then $\pdiff{f_i}{x_j}=A_{ij}$, $\pdiff{f_i}{u_j}=B_{ij}$, and $f_i(0,0)=c_i$.
\end{proof}
\begin{proposition}\label{prop:lin_pos_equiv}
System \eqref{eq:transformed_linear} is positive if and only if \eqref{eq:sign_cond_A}--\eqref{eq:sign_cond_c} hold with $S_{ij}^\rmstate=\{1,0\}$, $S_{ij}^\rminput=\{1,0\}$, and $S_i^\rmzero=\{1,0\}$.
\end{proposition}
\begin{proof}
Sufficiency follows from the nonlinear case. For necessity, if any entry violates the stated sign condition, one can construct nonnegative $x$ and $u$ that yield a negative component in $f^\rmc$ by Proposition~\ref{prop:nonlin_pos_f_cont} or in $f^\rmd$ by Proposition~\ref{prop:nonlin_pos_f_disc}, contradicting positivity.
\end{proof}
\subsection{Sign constraints for exactly linearizable models}
For EL models, the following theorem shows that the proposed transformation structure reduces sign-constraint verification to the signs of $A,B,c$.

\begin{theorem}\label{thm:exlin_sign_cond}
Assume $\Phi$ is componentwise strictly increasing.
Then:
\begin{enumerate}
    \item If $\Psi$ is independent of $x$, systems \eqref{eq:standard_el_ct}--\eqref{eq:standard_el_dt} satisfy the state sign constraints determined by $S_{ij}^\rmstate$ if and only if \eqref{eq:sign_cond_A} holds.
    \item If $\Psi$ is componentwise strictly increasing in $u$, systems \eqref{eq:standard_el_ct}--\eqref{eq:standard_el_dt} satisfy the input sign constraints determined by $S_{ij}^\rminput$ if and only if \eqref{eq:sign_cond_B} holds.
    \item If $\Phi(0)=0$ and $\Psi(0;0)=0$, systems \eqref{eq:standard_el_ct}--\eqref{eq:standard_el_dt} satisfy the origin sign constraints determined by $S_i^\rmzero$ if and only if \eqref{eq:sign_cond_c} holds.
\end{enumerate}
\end{theorem}
\begin{proof}
Let $\zeta:=A\Phi(x)+B\Psi(u;x)+c$.
Then $f^\rmc(x,u):=\left(\pdiff{\Phi}{x}\right)^{-1}\zeta$ and $f^\rmd(x,u):=\Phi^{-1}(\zeta)$.

(1) Using the chain rule and assumptions on $\Phi,\Psi$,
\begin{align}
    \pdiff{f_i^\rmc}{x_j}
    &=\left(\pdiff{\Phi_i}{x_i}\right)^{-1}A_{ij}\pdiff{\Phi_j}{x_j}
    -\frac{\frac{\del^2\Phi_i}{\del x_i\del x_j}}{\left(\pdiff{\Phi_i}{x_i}\right)^2}\zeta_i, \label{eq:proof_ct_state}\\
    \pdiff{f_i^\rmd}{x_j}
    &=\left.\pdiff{\Phi_i^{-1}}{\xi_i}\right|_{\xi_i=\zeta_i}
    A_{ij}\pdiff{\Phi_j}{x_j}. \label{eq:proof_dt_state}
\end{align}
Because $\Phi$ is componentwise strictly increasing, $\pdiff{\Phi_i}{x_i}>0$, $\pdiff{\Phi_i^{-1}}{\xi_i}>0$, and $\del^2\Phi_i/(\del x_i\del x_j)=0$ for $i\neq j$; hence $\sigma(\pdiff{f_i}{x_j})=\sigma(A_{ij})$.

(2) Similarly,
\begin{align}
\pdiff{f_i^\rmc}{u_j}
&=\left(\pdiff{\Phi_i}{x_i}\right)^{-1}B_{ij}\pdiff{\Psi_j}{u_j}, \\
\pdiff{f_i^\rmd}{u_j}
&=\left.\pdiff{\Phi_i^{-1}}{\xi_i}\right|_{\xi_i=\zeta_i}
B_{ij}\pdiff{\Psi_j}{u_j}.
\end{align}
Since $\pdiff{\Phi_i}{x_i}>0$, $\pdiff{\Phi_i^{-1}}{\xi_i}>0$, and $\pdiff{\Psi_j}{u_j}>0$, we have $\sigma(\pdiff{f_i}{u_j})=\sigma(B_{ij})$.

(3) With $\Phi(0)=0$ and $\Psi(0;0)=0$, $f_i^\rmc(0,0)=\left(\pdiff{\Phi_i}{x_i}\big|_{x=0}\right)^{-1}c_i$ and $f_i^\rmd(0,0)=\Phi_i^{-1}(c_i)$.
Componentwise strict increase yields $\sigma(f_i(0,0))=\sigma(c_i)$.
\end{proof}
\begin{theorem}\label{thm:exlin_pos_cond}
Assume $\Phi$ is componentwise strictly increasing, $\Psi$ is componentwise strictly increasing in $u$, $\Phi(0)=0$, and $\Psi(0;x)=0$ for all $x\in\bbR^{n_x}$.
Then \eqref{eq:standard_el_ct}--\eqref{eq:standard_el_dt} are positive if and only if \eqref{eq:sign_cond_A}--\eqref{eq:sign_cond_c} hold with $S_{ij}^\rmstate=\{1,0\}$, $S_{ij}^\rminput=\{1,0\}$, and $S_i^\rmzero=\{1,0\}$.
\end{theorem}
\begin{proof}
Under these assumptions, $x\in\bbR^{n_x}_\geq$ if and only if $\Phi(x)\in\bbR^{n_x}_\geq$, and $u\in\bbR^{n_u}_\geq$ if and only if $\Psi(u;x)\in\bbR^{n_u}_\geq$.
Thus, positivity of the EL models is equivalent to positivity of \eqref{eq:transformed_linear}; Proposition~\ref{prop:lin_pos_equiv} provides the conditions.
\end{proof}
\section{Neural Parameterization and Convex MPC}
\subsection{Neural parameterization of sign-constrained EL}
We represent the input transformation as
\begin{align}
    \Psi(u;x)&=\varphi_\Psi^{(L_\Psi)}\big(\cdots\varphi_\Psi^{(2)}(\varphi_\Psi^{(1)}(u;x);x);x\big), \label{eq:psi_param} \\
    \varphi_\Psi^{(i)}(u;x)
    &=\sinh^{-1}\!\big(a_\Psi^{(i)}(x)+\sinh(\nonumber\\
    &\quad\quad\quad\quad\quad W_\Psi^{(i)}(x)\odot u+b_\Psi^{(i)}(x))\big)\nonumber\\
    &\quad -\sinh^{-1}\!\big(a_\Psi^{(i)}(x)+\sinh(b_\Psi^{(i)}(x))\big). 
\end{align}
Here $i\in\bbZ_{L_\Psi}$.
The functions $a_\Psi^{(i)}(x)$, $b_\Psi^{(i)}(x)$, and $W_\Psi^{(i)}(x)$ are parameterized by neural networks in $x$.
This follows \cite{exlin}, except that the constant term is adjusted so that $\Psi(0;x)=0$.
The state transformation $\Phi$ is parameterized analogously with $a_\Phi^{(i)},b_\Phi^{(i)},W_\Phi^{(i)}$ and with the same zero-origin adjustment.
To enforce componentwise strict increase, let ${W_\Psi'}^{(i)}(x)$ be free functions and ${W_\Phi'}^{(i)}$ be free parameters, and set $W_\Psi^{(i)}(x)=\operatorname{ReLU}\left({W_\Psi'}^{(i)}(x)\right)+\varepsilon$ and $W_\Phi^{(i)}=\operatorname{ReLU}\left({W_\Phi'}^{(i)}\right)+\varepsilon$ with $\varepsilon>0$.

For coefficient signs, define the map from a free scalar $\theta$ to a constrained scalar by
\begin{align}
    \pi_{\{s\}}(\theta)&:=s\left(\operatorname{ReLU}(\theta)+\varepsilon\right), 
    &&s\in\{1,-1,0\}, \label{eq:pi_single}\\
    \pi_{\{s,0\}}(\theta)&:=s\operatorname{ReLU}(\theta), 
    &&s\in\{1,-1\}, \label{eq:pi_two}\\
    \pi_{\{1,-1,0\}}(\theta)&:=\theta . \label{eq:pi_free}
\end{align}
Then
\begin{align}
    A_{ij}&=\pi_{S_{ij}^\rmstate}(A'_{ij}),\quad
    B_{ij}=\pi_{S_{ij}^\rminput}(B'_{ij}),\nonumber\\
    c_i&=\pi_{S_i^\rmzero}(c'_i). \label{eq:matrix_param}
\end{align}
where the primed variables are free parameters.
Equations \eqref{eq:psi_param}--\eqref{eq:matrix_param} enforce the assumptions of Theorem~\ref{thm:exlin_sign_cond} by construction.
In particular, with $S_i^\rmzero=\{0\}$ for all $i$, \eqref{eq:matrix_param} gives $c=0$, which is the same zero-origin equilibrium preservation as in Remark~\ref{rem:equilibrium_preservation}.

\subsection{Convex formulation of constrained tracking MPC}
Consider the original-coordinate optimal control formulation
\begin{align}
    \min_{u}\sum_{t=0}^{N-1}&w^\req\|x(t)-x^\req(t)\|^2+w^u\|u(t)\|^2\nonumber\\
    &+w^\soft\|\max(x(t)-\overline{x}^\soft(t),0)\|^2 \label{eq:mpc_original_obj}\\
    \st\quad
    &x(t+1)=\Phi^{-1}\big(\eta(t)\big), \label{eq:mpc_original_dyn}\\
    &\eta(t)=A\Phi(x(t))+B\Psi(u(t);x(t))+c, \nonumber\\
    &0\leq x(t),\quad 0\leq u(t), \label{eq:mpc_original_pos}\\
    &\underline{x}(t)\leq x(t)\leq\overline{x}(t),\quad
    \underline{u}(t)\leq u(t)\leq\overline{u}(t), \label{eq:mpc_original_bounds}\\
    &t=0,\ldots,N-1. \nonumber
\end{align}
Given the current state $x$, set $x(0)=x$, solve \eqref{eq:mpc_original_obj}--\eqref{eq:mpc_original_bounds}, and apply the first optimal input $u^*(0)$.

Using $\xi=\Phi(x)$, $v=\Psi(u;x)$, and slack variables for the soft constraints, we reformulate constrained tracking in transformed coordinates as
\begin{align}
    \min_{v,\gamma^\soft}\sum_{t=0}^{N-1}&w^\req\|\xi(t)-\Phi(x^\req(t))\|^2
    +w^v\|v(t)\|^2\nonumber\\
    &+w^\soft\|\gamma^\soft(t)\|^2 \label{eq:mpc_transformed_obj}\\
    \st\quad
    &\xi(t+1)=A\xi(t)+Bv(t)+c, \label{eq:mpc_transformed_dyn}\\
    &0\leq\xi(t),\quad0\leq v(t), \label{eq:mpc_transformed_pos}\\
    &\Phi(\underline{x}(t))\leq\xi(t)\leq\Phi(\overline{x}(t)), \label{eq:mpc_transformed_x_bounds}\\
    &\Psi(\underline{u}(t);\Phi^{-1}(\xi(t)))\leq v(t), \nonumber\\
    &v(t)\leq\Psi(\overline{u}(t);\Phi^{-1}(\xi(t))), \label{eq:mpc_input_bound_v}\\
    &0\leq\gamma^\soft(t),\quad
    \xi(t)-\Phi(\overline{x}^\soft(t))\leq\gamma^\soft(t), \label{eq:mpc_transformed_soft}\\
    &t=0,\ldots,N-1. \nonumber
\end{align}
If $\Psi$ is independent of $x$, the optimization problem defined by \eqref{eq:mpc_transformed_obj}--\eqref{eq:mpc_transformed_soft} is a convex QP.
Sign constraints themselves are not the direct source of convexity; the componentwise increasing and zero-origin structure that enables sign constraints also makes the constraints convex in transformed coordinates.
For a state-dependent $\Psi$, \eqref{eq:mpc_input_bound_v} is generally nonconvex for $t\geq1$.
We use a first-step input-constraint relaxation: the input bounds in \eqref{eq:mpc_input_bound_v} are enforced only for $t=0$, whereas all other constraints are retained over the horizon.
This relaxation loses exact equivalence to \eqref{eq:mpc_original_obj}--\eqref{eq:mpc_original_bounds} and the input-bound guarantee for planned future inputs at $t\geq1$, but it preserves the input constraint applied at the current step and yields a convex QP.

\begin{proposition}\label{prop:mpc_qp_properties}
Suppose the weights in \eqref{eq:mpc_transformed_obj} are positive, the transformed-coordinate QP \eqref{eq:mpc_transformed_obj}--\eqref{eq:mpc_transformed_soft} or its first-step input-constraint relaxation is feasible, and the linear independence constraint qualification holds at the optimal solution for the given initial state.
Then the solution is unique, globally optimal, and locally Lipschitz continuous with respect to the initial state.
\end{proposition}
The result follows from strong convexity of the QP objective and standard regularity properties of optimization-based controllers \cite{regularity_opt}.
Convex QPs also offer practical advantages: their computational load scales polynomially with dimension and horizon length \cite{boyd_convex}, and mature solvers return solutions without relying on a favorable initial guess.

\begin{remark}
The formulation can be extended beyond the quadratic tracking cost in \eqref{eq:mpc_transformed_obj} to any cost that is convex in transformed coordinates, including tracking costs based on any norm and linear economic costs.
\end{remark}
\section{Numerical Examples}
\subsection{Example 1: Three-tank system}
The first example is a benchmark three-tank system with monotonicity and positivity constraints.
Consider
\begin{align}
    \dot{h}_1&=-k_1\sqrt{h_1}+k_6V_{\rmp,1}, \nonumber\\
    \dot{h}_2&=k_2\sqrt{h_1}-k_3\sqrt{h_2}+k_7V_{\rmp,2}, \label{eq:three_tank}\\
    \dot{h}_3&=k_4\sqrt{h_2}-k_5\sqrt{h_3}. \nonumber
\end{align}
Let $x=[h_1,h_2,h_3]$ and $u=[V_{\rmp,1},V_{\rmp,2}]$.
For the discrete-time model,
\begin{align}
    S^\rmstate&=\left[\begin{array}{@{}ccc@{}}
    \{1,0\}&\{0\}&\{0\}\\
    \{1,0\}&\{1,0\}&\{0\}\\
    \{0\}&\{1,0\}&\{1,0\}
    \end{array}\right], \nonumber\\
    S^\rminput&=\left[\begin{array}{@{}cc@{}}
    \{1,0\}&\{0\}\\
    \{0\}&\{1,0\}\\
    \{0\}&\{0\}
    \end{array}\right],\quad
    S^\rmzero=\left[\begin{array}{@{}c@{}}\{0\}\\\{0\}\\\{0\}\end{array}\right]. \label{eq:three_tank_sign}
\end{align}
These signs express downstream propagation, nonnegative pump effects on the directly actuated tanks, and the zero-input equilibrium at the origin.
Here, $\Psi$ is taken independent of the state, so the control formulation requires no relaxation.
Because the system is monotone, we also use an ICNN-based controller as a baseline.

Data are generated by integrating \eqref{eq:three_tank} using LSODA \cite{odepack} with $k_1=0.020$, $k_2=0.009$, $k_3=0.015$, $k_4=0.010$, $k_5=0.013$, $k_6=0.0025$, and $k_7=0.0020$.
We use 1000 trajectories with a sampling time of $1.0$ s and a duration of $200$ s, initial levels in $[0,0.50]$ m, and pump voltages in $[0,9.0]$ V.
The data are split as follows: 70\% training, 15\% validation, and 15\% testing.
For interpolation, the split is random; for extrapolation, trajectories are sorted by the minimum water level before splitting.
Training uses $\varepsilon=10^{-8}$, a learning rate of $5\times10^{-4}$, a batch size of 512, and an early-stopping patience of 15.
For EL models, $L_\Psi,L_\Phi\in\{2,3,4\}$ are tested; for NN and ICNN baselines, hidden layers and hidden units are each selected from $\{2,3,4\}$.

The original-coordinate control formulation is the specialization of \eqref{eq:mpc_original_obj}--\eqref{eq:mpc_original_bounds} given by
\begin{align}
    \min_{u}\sum_{t=0}^{N-1}&w^\req[h_3(t)-h_3^\req(t)]^2
    +w^u\|u(t)\|^2 \nonumber\\
    \st\quad &\eqref{eq:mpc_original_dyn},\quad u(t)\geq0,
    \quad t=0,\ldots,N-1. \label{eq:tank_mpc}
\end{align}
For MPC, we use the extrapolation-best model with $N=10$, a total control duration of $50$ s, $x(0)=[0.25,0.25,0.25]$ m, $h_3^\req(t)=0.24$ m, $w^\req=1.0$, and $w^u=10^{-5}$.

\subsection{Example 2: Hybrid powertrain}
\begin{figure}[t]
    \centering
    \includegraphics[width=0.99\linewidth]{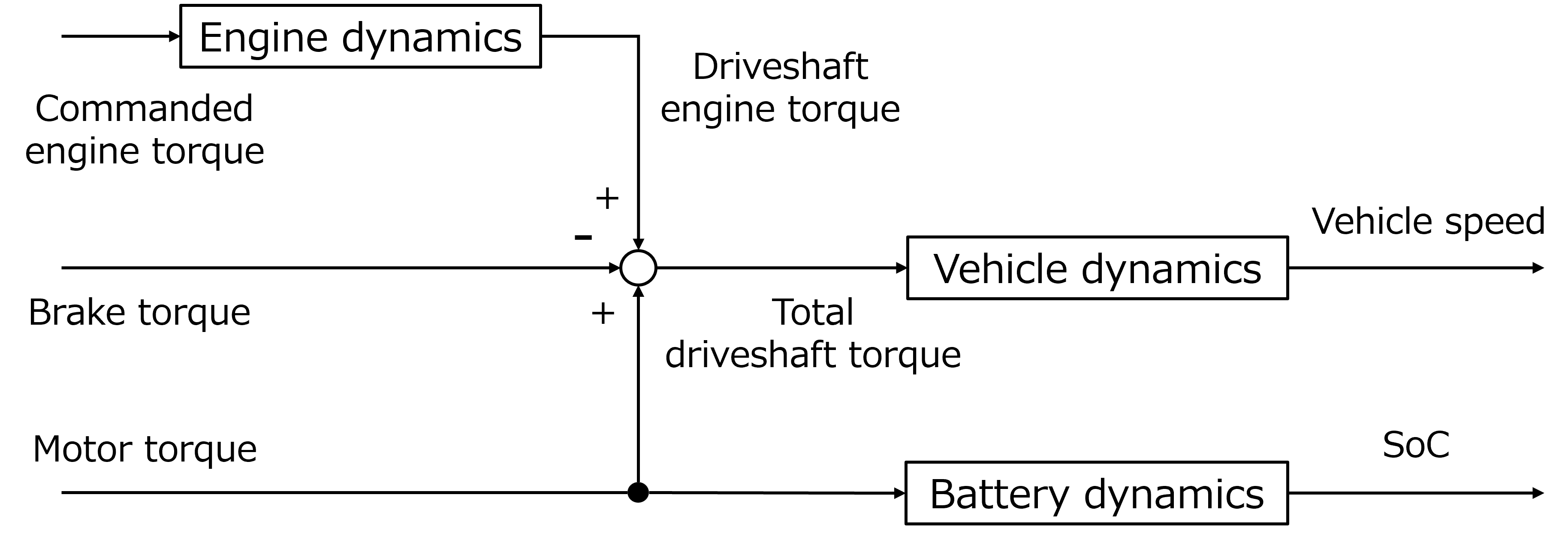}
    \caption{Hybrid powertrain system.}
    \label{fig:hev}
\end{figure}

The second example is a hybrid powertrain as an engineering application with mixed positive, negative, and zero input signs.
Fig.~\ref{fig:hev} shows the hybrid powertrain system.
Let $x=[\tau^{\eng,\rmds},V,S]$, where $\tau^{\eng,\rmds}$ is the net driveshaft engine torque after friction losses and can be negative, $V$ is the vehicle speed, and $S$ is the state of charge (SoC).
Let $u=[\tau^{\eng,\cmd},\tau^\mot,\tau^\brk]$.
The sign constraints are
\begin{align}
    S_{ij}^\rmstate&=\{1,-1,0\}\quad (i,j\in\bbZ_3), \nonumber\\
    S^\rminput&=\left[\begin{array}{@{}ccc@{}}
        \{1,0\}&\{0\}&\{0\}\\
        \{1,0\}&\{1,0\}&\{-1,0\}\\
        \{0\}&\{-1,0\}&\{0\}
    \end{array}\right], \quad
    S^\rmzero=\left[\begin{array}{@{}c@{}}
        \{0\}\\\{0\}\\\{0\}
    \end{array}\right]. \label{eq:hev_sign}
\end{align}
They encode that commanded engine torque increases net engine torque and speed, motor torque increases speed and decreases SoC, brake torque decreases speed, and the origin is an equilibrium.
For this example, $\Psi$ is allowed to depend on the state, and the first-step input-constraint relaxation is used in MPC.
An ICNN is not used because the dynamics are not monotone.

The original-coordinate control formulation is
\begin{align}
    &\hspace{-2mm}\min_{u}\sum_{t=0}^{N-1}
    w^\req[V(t)-V^\req(t)]^2+w^\eng[\tau^{\eng,\cmd}(t)]^2 \nonumber\\
    &+w^\mot[\tau^\mot(t)]^2+w^\brk[\tau^\brk(t)]^2 \nonumber\\
    &+w^\soc\Big[\max\big(S^\target(V^\req(t))
    -S(t),0\big)\Big]^2 \nonumber\\
    &\hspace{-2mm}\st\quad \eqref{eq:mpc_original_dyn},\quad
    0\leq\tau^{\eng,\cmd}(t),\quad 0\leq\tau^\brk(t), \nonumber\\
    &\tau^{\eng,\rmds}(t)\leq\overline{\tau}^{\eng,\rmds}(V^\req(t)),\quad
    \underline{S}\leq S(t)\leq\overline{S}, \nonumber\\
    &\underline{\tau}^\mot\leq\tau^\mot(t)\leq\overline{\tau}^\mot, \nonumber\\
    &\frac{\underline{P}^\mot}{V^\req(t)}\leq\tau^\mot(t)\leq
    \frac{\overline{P}^\mot}{V^\req(t)},\quad
    \tau^\brk(t)\leq\overline{\tau}^\brk \label{eq:hev_mpc_const} \\
    & t=0,\ldots,N-1 . \nonumber
\end{align}
Here $S^\target(V):=S^\rmref-S^\reg(V)$, where $S^\reg(V)$ denotes the recoverable SoC until standstill and is obtained from a calibrated map.
The driveshaft engine-torque upper bound $\overline{\tau}^{\eng,\rmds}(V)$ is also obtained through calibrated maps, based on the regulatory NOx bound.

Powertrain Blockset (MathWorks) is used to generate trajectories with trapezoidal commanded torques: $\tau^{\eng,\cmd}\in[0,198]$ Nm, $\tau^\mot\in[-198,196]$ Nm, and $\tau^\brk\in[0,442]$ Nm.
The dataset contains 2583 trajectories with a sampling time of $0.1$ s and a duration of $100$ s. Only data for which the engine speed lies in $[0,5000]$ rpm are retained.
The interpolation and extrapolation splits follow the same 70/15/15 rule as the tank example, sorting by the minimum net driveshaft engine torque for extrapolation.
Training uses $\varepsilon=10^{-8}$, a learning rate of $10^{-4}$, a batch size of 1024, and an early-stopping patience of 10.
For EL models, $L_\Psi,L_\Phi$, and the layer count of the $a,b,W$ networks are selected from $\{2,3,4\}$, and the hidden size from $\{2,3,4\}$; for the NN baseline, hidden layers and hidden units are selected from $\{4,5,6\}$.
For MPC, the extrapolation-best model is used with $N=10$, a total duration of $10$ s, $x(0)=[10\ \mathrm{Nm},60\ \mathrm{km/h},30\ \mathrm{Ah}]$, $S^\rmref=30$, $\underline{S}=20$, $\overline{S}=40$, $\underline{\tau}^\mot=-200$, $\overline{\tau}^\mot=200$, $\underline{P}^\mot=-10000$, $\overline{P}^\mot=10000$, $\overline{\tau}^\brk=50$, and weights $w^\req=10$, $w^\eng=w^\mot=0.001$, $w^\brk=0.01$, $w^\soc=0.5$.

\subsection{Results}
\begin{figure*}[t]
    \centering
    \begin{minipage}{0.47\textwidth}
        \centering
        \includegraphics[width=0.8\linewidth]{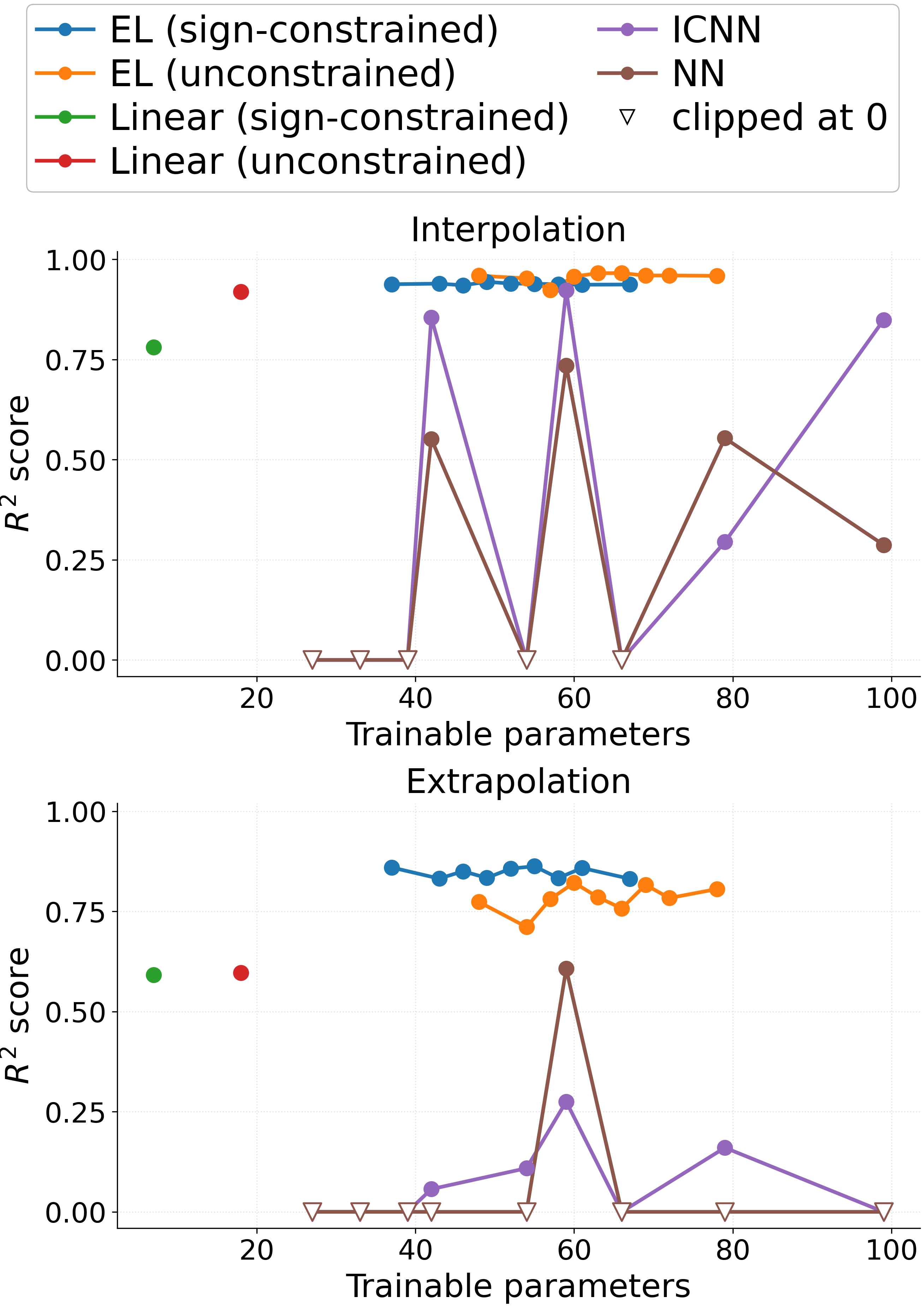}
    \end{minipage}\hfill
    \begin{minipage}{0.47\textwidth}
        \centering
        \includegraphics[width=0.93\linewidth]{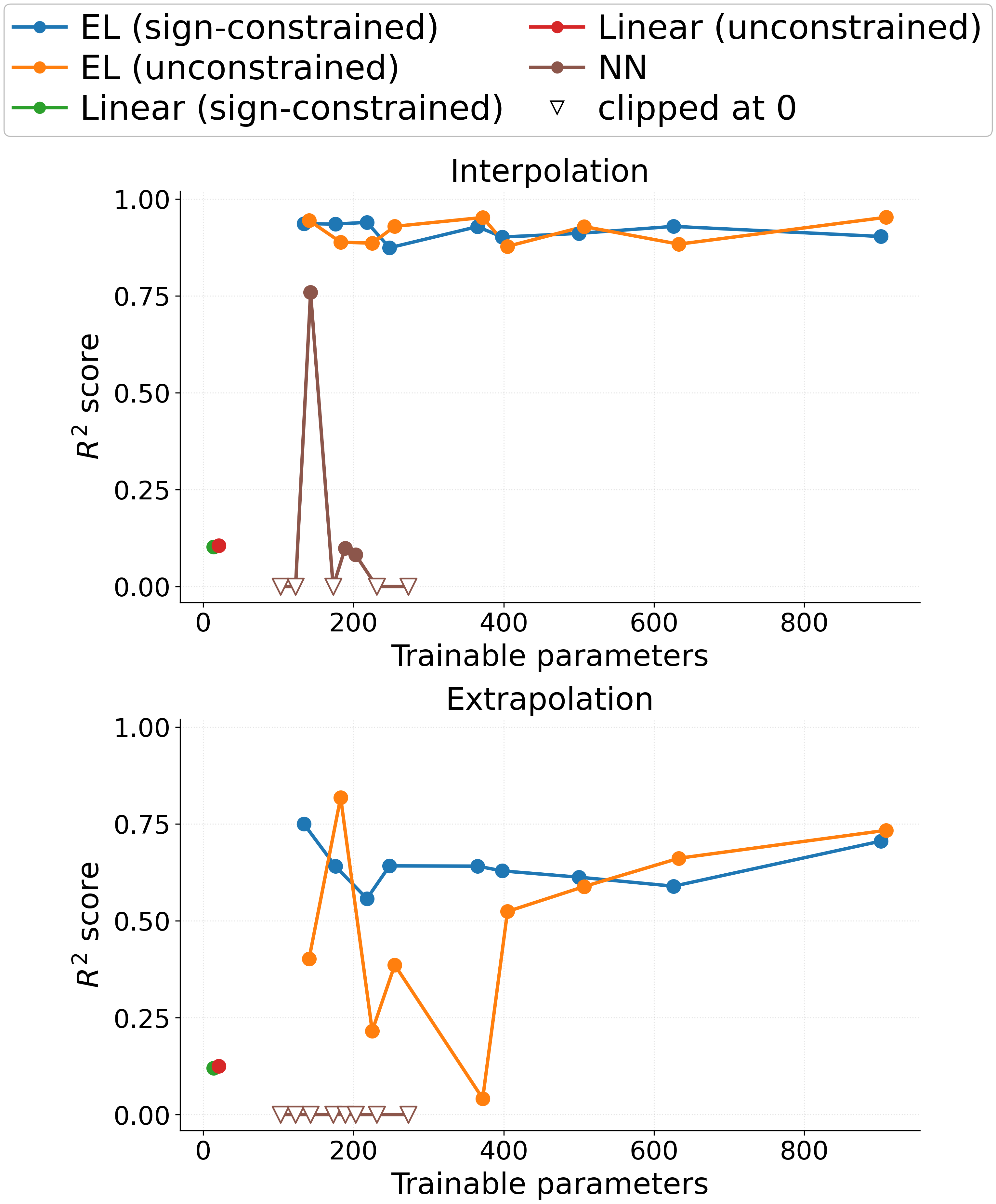}
    \end{minipage}
    \caption{$R^2$ scores for one-step state changes under different parameter counts. Left: three-tank system. Right: hybrid powertrain.
    Negative scores are clipped at $0$ for visualization and shown by triangular markers.
    }
    \label{fig:r2_results}
\end{figure*}

\begin{figure*}[t]
    \centering
    \includegraphics[width=0.98\textwidth]{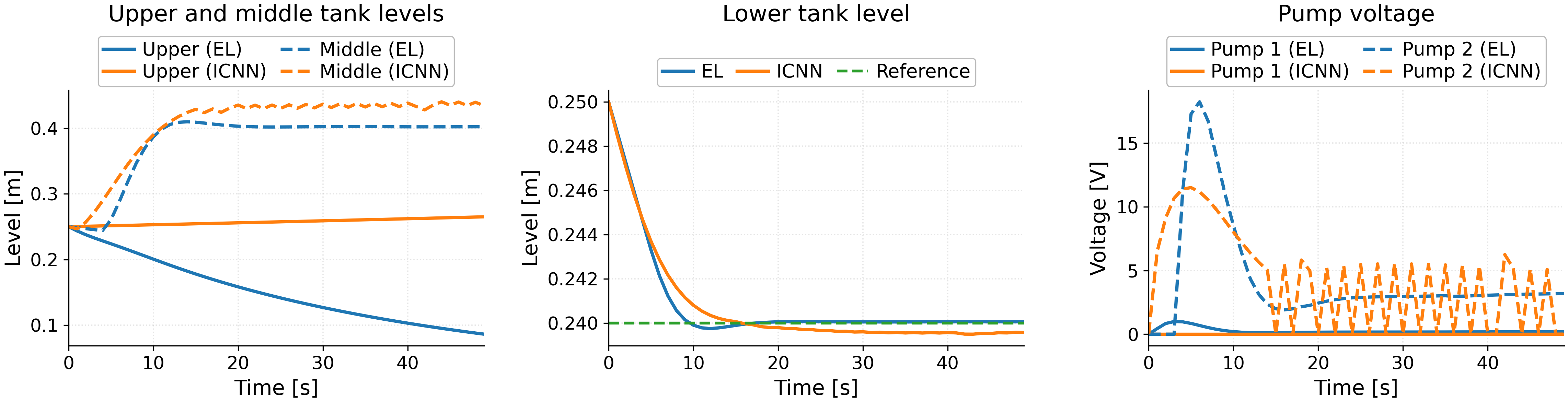}
    \caption{MPC trajectories in the three-tank system.}
    \label{fig:mpc_results_tank}
\end{figure*}

\begin{figure*}[t]
    \centering
    \includegraphics[width=0.98\textwidth]{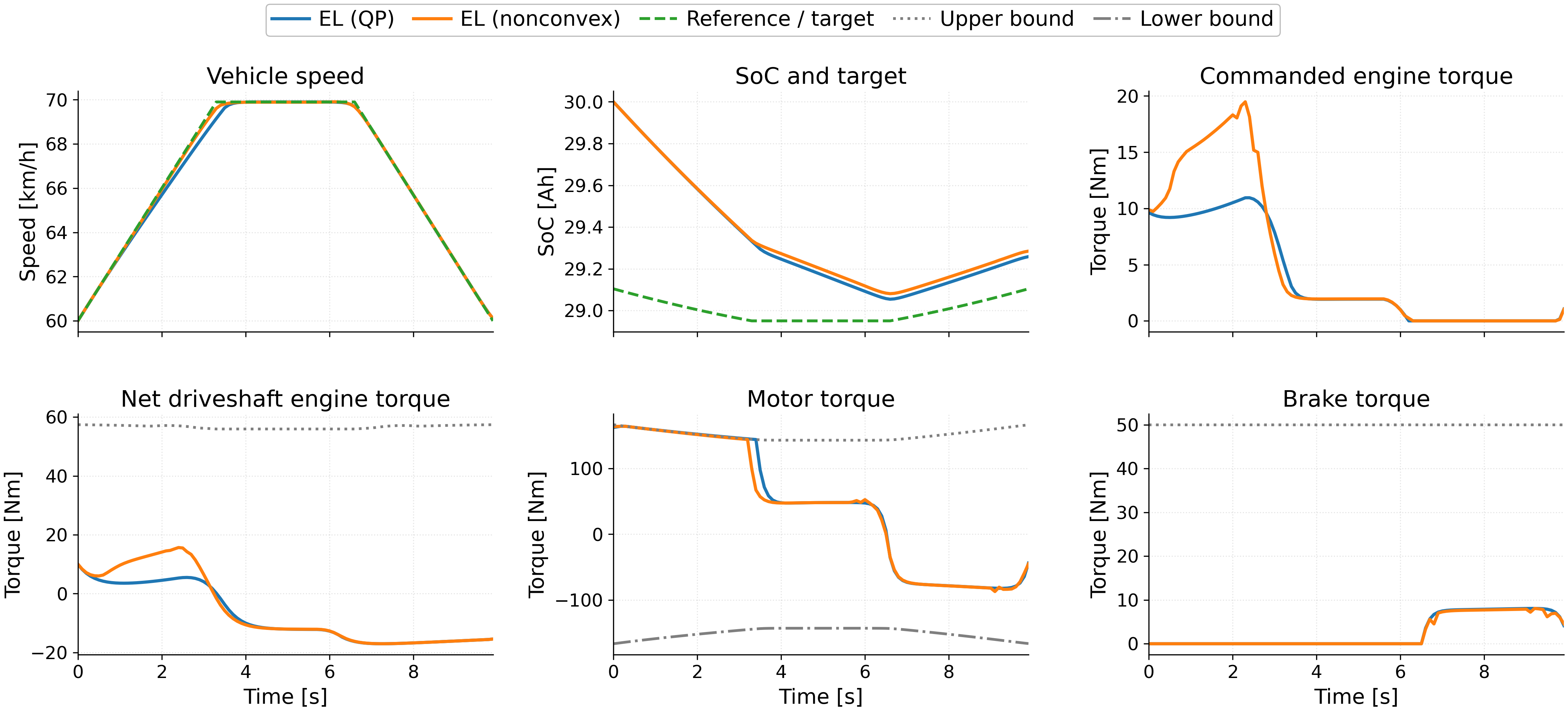}
    \caption{MPC trajectories in the hybrid powertrain. The SoC panel shows $S^\target$ rather than $S^\rmref$.}
    \label{fig:mpc_results_hev}
\end{figure*}

Fig.~\ref{fig:r2_results} shows one-step prediction accuracy over model sizes.
In both examples, sign-constrained EL models maintain high extrapolation accuracy over a broad parameter range, whereas the unconstrained NN is more sensitive to hyperparameters and degrades severely outside the training region.
Table~\ref{tab:mpc_evaluation} and Figs.~\ref{fig:mpc_results_tank}--\ref{fig:mpc_results_hev} summarize the control results.
\begin{table*}[t]
    \centering
    \caption{MPC evaluation. Values are mean $\pm$ standard deviation.}
    \renewcommand{\arraystretch}{1.15}
    \setlength{\tabcolsep}{5pt}
    \scriptsize
    \begin{tabular}{|c|c|c|c|c|}
        \hline
        Example & Method & Optimizer & Input variation norm & Optimization time [s] \\
        \hline
        Three-tank & Sign-constrained EL (convex) & OSQP \cite{osqp} & $0.7541\ (\pm 1.8730)$ & $0.0409\ (\pm 0.0089)$ \\
        \hline
        Three-tank & ICNN (nonconvex) & SLSQP \cite{slsqp_orig} & $3.8185\ (\pm 2.2501)$ & $0.0325\ (\pm 0.0199)$ \\
        \hline
        Hybrid powertrain & Sign-constrained EL (convex relaxation) & OSQP \cite{osqp} & $2.9772\ (\pm 7.5532)$ & $0.0714\ (\pm 0.0102)$ \\
        \hline
        Hybrid powertrain & Sign-constrained EL (without relaxation) & SLSQP \cite{slsqp_orig} & $3.3846\ (\pm 7.2642)$ & $36.1299\ (\pm 13.1739)$ \\
        \hline
    \end{tabular}
    \label{tab:mpc_evaluation}
\end{table*}
In the three-tank case, the convex QP produces much lower input variation than the ICNN-based nonconvex formulation.
In the hybrid powertrain case, the convex relaxation yields input variation similar to that of the nonconvex formulation while reducing solve time by orders of magnitude and avoiding large solve-time variability.

\section{Conclusion}
We introduced sign constraints that generalize monotonicity, positivity, and sign-definiteness, and proposed EL deep models that satisfy these constraints by construction.
The same structure enables constrained tracking MPC to be formulated as a convex QP or a convex relaxation.
On a three-tank system and a hybrid powertrain, the proposed models improved extrapolation performance and produced smooth control inputs with efficient optimization.
Future work will focus on developing tighter convex relaxations for state-dependent input transformations and analyzing the resulting optimality gaps theoretically.
Other directions include automatic specification of sign constraints, robust and stochastic extensions, online model adaptation, and experimental validation.

\begin{ack}
We thank Hideto Inagaki and Matsuei Ueda for valuable discussions.
\end{ack}

\bibliographystyle{unsrt}
\bibliography{references}

\end{document}